\documentclass[11pt, A4]{article}
\usepackage{overpic}
\usepackage{graphicx}
\usepackage{epstopdf}
\usepackage{amsmath}
\usepackage{amssymb}
\usepackage{amsfonts}
\usepackage{amsthm}
\usepackage[usenames]{color}
\usepackage{array}
\usepackage{subfigure}
\definecolor{AV}{rgb}{0.65,0.0,0}
\definecolor{GC}{rgb}{0,0.0,0.65}

\usepackage[
      colorlinks=true,
      linkcolor=blue,
      urlcolor=blue,
      filecolor=blue,
      citecolor=red,
      pdfstartview=FitV,
      pdftitle={},
      pdfauthor={},
      pdfsubject={},
      pdfkeywords={},
      pdfpagemode=None,
      bookmarksopen=true
]{hyperref}

\usepackage{epsfig}

\usepackage{hyperref}

\def\cO{\mathcal{O}}

\def\cQ{\mathcal{Q}}

\newcommand{\Sg}{\Sigma}

\newcommand{\tht}{\bar{\theta}}

\def\half{\frac{1}{2}}

\def\beq{\begin{eqnarray}}
\def\eeq{\end{eqnarray}}

\def\mf{\mathfrak}
\def\be{\begin{equation}}
\def\ee{\end{equation}}
\def\bea{\begin{eqnarray}}
\def\eea{\end{eqnarray}}

\newcommand{\rom}[1]{\mathrm{#1}}

\def\cO{\mathcal{O}}

\def\cQ{\mathcal{Q}}

\def\mf{\mathfrak}

\def\nn{\nonumber}
\def\half{\frac{1}{2}}

\numberwithin{equation}{section}

\begin{document}
\pagestyle{myheadings}
\markboth{\textsc{\small }}{%
  \textsc{\small Extremal limits of the Cveti\v c-Youm black hole and
nilpotent orbits of $\mathrm{G}_{2(2)}$ }} \addtolength{\headsep}{4pt}

%% \begin{titlepage}

\begin{flushright}
\texttt{ULB-TH/10-27}
\end{flushright}

\begin{centering}

  \vspace{0cm}

  \textbf{\Large{Extremal limits of the Cveti\v c-Youm black hole \\ and
nilpotent orbits of $\mathrm{G}_{2(2)}$ }}

  \vspace{0.8cm}

  {\large Josef Lindman H\"ornlund and Amitabh Virmani}

  \vspace{0.5cm}

\begin{minipage}{.9\textwidth}\small \it \begin{center}
  Physique Th\'eorique et Math\'ematique,  \\ Universit\'e Libre de
    Bruxelles and International Solvay Institutes\\ Campus
    Plaine C.P. 231, B-1050 Bruxelles,  Belgium\\
   {\tt jlindman, avirmani@ulb.ac.be}
    \end{center}
\end{minipage}

\end{centering}

\vspace{1cm}

%\begin{center}
%  \begin{minipage}{.9\textwidth}
%    \begin{center}

\begin{abstract}
We study extremal cohomogeneity one five-dimensional asymptotically flat black holes of minimal supergravity in terms of the geodesics generated  by nilpotent elements of the Lie algebra $\mf{g}_{2(2)}$ on the coset manifold $\mathrm{G}_{2(2)}/\mathrm{SO}(2,2)$. There are two branches of regular extremal black holes with these properties: (i) the supersymmetric BMPV branch, and (ii) the non-supersymmetric extremal branch. We show that both of these branches are reproduced by nilpotent $\mathrm{SO}(2,2)$-orbits. Furthermore, we show that the partial ordering of nilpotent orbits of $\mathrm{G}_{2(2)}$ is in one-to-one correspondence with the phase diagram of these extremal black holes.
\end{abstract}
\vfill

\noindent \mbox{}
\raisebox{-3\baselineskip}{%
  \parbox{\textwidth}{ \mbox{}\hrulefill\\[-4pt]}}

\thispagestyle{empty} \newpage

\tableofcontents

\setcounter{equation}{0}

\section{Introduction}
Asymptotically flat single center black holes of various four-dimensional supergravity theories have been extensively studied in the literature. At the classical level many physical and thermodynamical properties of these black holes are  well understood. For a large number of theories single center four-dimensional black holes for any allowed configuration of charges can be explicitly constructed. This is particularly so for extremal black holes---BPS as well as non-BPS---where in a large number of cases a complete classification is already available. See for instance \cite{Ceresole:2007wx, Andrianopoli:2007gt, Hotta:2007wz, Gaiotto:2007ag, Gimon:2007mh, Bergshoeff:2008be, Bellucci:2008sv, Perz:2008kh,  Bossard:2009at,  Compere:2009zh, Chemissany:2009hq, Bellucci:2009qv, Bossard:2009mz, Gunaydin:2009pk, Bossard:2009we, Ferrara:2009bw, Kim:2010bf,   Meessen:2010fh, Chemissany:2010zp, Compere:2010fm} for recent discussions and references therein. Algebraic methods, see e.g., \cite{Bossard:2009at, Compere:2009zh, Bossard:2009mz, Gunaydin:2009pk, Bossard:2009we, Kim:2010bf}, in terms of a three dimensional sigma-model with symmetric target space $G/\tilde{K}$ have proven to be extremely useful in this regard. These methods provide us with a systematic technique to find all extremal as well as non-extremal black holes. See \cite{Compere:2009zh, Kim:2010bf, Compere:2010fm} for detailed explicit calculations for the $N=2,~D=4~$S$^3$ model. (S$^3$ model is obtained by setting S=T=U in the STU model.) Practical implementation of these methods may be technically challenging for supergravity theories of interest, but, there are no conceptual obstacles as to the applicability of these techniques. These developments have led to, among other things, a fairly complete understanding of the attractor mechanism for BPS as well as extremal non-BPS black holes \cite{Bossard:2009we}.

It is natural to ask how much of these considerations extend to five-dimensional asymptotically flat black holes. Non-extremal five-dimensional black holes have been explored in various contexts from the three-dimensional sigma-model point of view \cite{Giusto:2007fx, Bouchareb:2007ax, Ford:2007th, Giusto:2007tt, Gal'tsov:2008nz, Camps:2008hb, Compere:2009zh, Gal'tsov:2009da}, but such studies have not yet been adapted to extremal black holes. Reference \cite{Berkooz:2008rj} took some steps in this direction. From the sigma-model perspective, the difference between extremal and non-extremal black holes arises due to the fact that the relevant orbits for extremal black holes are nilpotent $\tilde K$-orbits of the three-dimensional hidden symmetry group, where as for non-extremal black holes the relevant orbits are semi-simple ones. Berkooz and Pioline in reference \cite{Berkooz:2008rj} studied supersymmetric five-dimensional black holes from the sigma-model perspective. They explicitly verified nilpotency of the charge matrix for certain supersymmetric five-dimensional black holes. However, a detailed study of various nilpotent orbits and how they relate to the known phase diagrams of five-dimensional black holes in a supergravity theory has been largely missing in the literature. The primary purpose of this paper is to fill this gap in the special case of minimal supergravity.

The main motivation for this study is to develop algebraic tools that can let us find new black hole solutions of physical interest. Many such solutions are still not known, for example,  a black ring that describes thermal excitations above the supersymmetric black ring is not known (it was conjectured to exist in \cite{Elvang:2004xi}; see \cite{Compere:2010fm} for a more detailed discussion), and certain dyonic extremal black rings, conjectured to exist in \cite{Kim:2010bf, Compere:2010fm}, are also not known. Another motivation is to understand how different nilpotent orbits, and their partial orderings, are to be interpreted from the five-dimensional spacetime point of view. We also hope that such a study will give us further insights into the attractor mechanism for five-dimensional extremal black holes.

The rest of the paper is organized as follows. In section \ref{CY} we review salient features of the Cveti\v c-Youm family of metrics \cite{Cvetic:1996xz} and discuss its various extremal limits. In particular, we discuss two classes of regular cohomogeneity one extremal black holes contained in this family.  In section \ref{sec:geodesics} we present a general discussion of five-dimensional spacetimes with two commuting Killing vectors from the three-dimensional sigma-model perspective. We show that the asymptotic boundary conditions relevant for five-dimensional black holes naturally give rise to a reductive decomposition of the Lie algebra of the hidden symmetry group of the theory. We then use this decomposition to relate five-dimensional spacetime charges to the Lie algebra valued Noether charge matrix of the sigma-model.  In section \ref{sec:nilpotent} we show that the nilpotent orbits of G$_{2(2)}$ are in one-to-one correspondence with phases of regular extremal cohomogeneity one black holes of minimal supergravity obtained as limits of the Cveti\v c-Youm solution. We close with a brief discussion in section \ref{sec:disc}. Technical details are regelated to appendices.  In appendix \ref{app:CY} we recall the Cveti\v c-Youm solution.  In appendix \ref{app:liealgebradetails} we explicitly present the reductive decomposition of
$\mathfrak{g}_{2(2)}$ relevant for the five-dimensional asymptotically flat boundary conditions. We remark that this work is largely a continuation of \cite{Kim:2010bf}. For a thorough discussion of nilpotent orbits of $\mathrm{G}_{2(2)}$ we refer the reader to \cite{Kim:2010bf}.

\section{The Cveti\v c-Youm family and its extremal limits}

\label{CY}
In this section we review certain properties of the Cveti\v c-Youm family of metrics. These metrics are studied in detail in \cite{Cvetic:1997uw, Cvetic:1996xz, Giusto:2004id,  Jejjala:2005yu, Dias:2007nj} so we shall be brief. Our main interest is in its various extremal limits and their sigma-model description.

\subsection{Preliminaries}

The theory we are interested in is minimal ungauged supergravity in five dimensions. It contains a metric $g_5$ and a gauge potential $A$ whose field strength is $F= dA$. The bosonic part of the Lagrangian takes the form of Einstein-Maxwell theory with a Chern-Simons term,
\begin{equation}
\mathcal{L}_5 = R_5 \star_5  \mathbf{1} - \half\star_5  F \wedge F +\frac{1}{3\sqrt 3} F \wedge F \wedge A. \label{eqn:5dsugra}
\end{equation}
Upon dimensional reduction on a circle this theory becomes the so called $N=2$, $D=4$, S$^3$ supergravity, which was analyzed from a group theory perspective in detail in \cite{Kim:2010bf}. Four-dimensional asymptotically flat black holes of the $\mathrm{S}^3$ supergravity when uplifted to minimal supergravity describe asymptotically Kaluza-Klein black holes in five dimensions. In this paper we are interested in \emph{five-dimensional asymptotically flat boundary conditions} as opposed to the Kaluza-Klein boundary conditions explored in \cite{Kim:2010bf}.

We parameterize five-dimensional Minkowski metric as
\begin{equation}
\label{eqn:Minkowski}
ds_5^2(\mathbb{R}^{4,1}) = -dt^2+d r^2+\frac{r^2}{4}(d\theta^2+d\phi^2+ d\psi^2+ 2 \cos \theta d\psi d\phi),
\end{equation}
where $r$ is the  radial coordinate and $\theta, \phi$, and  $\psi$ are the standard Euler angles on the three-sphere with
\be
0 \le \theta < \pi, \qquad 0 \le \phi < 2 \pi, \qquad 0 \le \psi < 4 \pi.
\ee
The action of $\partial_{\psi}$ and $\partial_{\phi}$ as Killing fields commute, but they are not orthogonal as vectors in general. In this paper we exclusively work with metrics with three commuting Killing symmetries generating an $\mathbb{R}$ $\times$ U(1) $\times$ U(1) isometry group. However, to have a three-dimensional picture of black holes as geodesics on a coset manifold, this isometry group is not sufficiently large. One must look at metrics where one of the U(1)'s is enhanced to an $\mathrm{SU}(2)$ symmetry. This happens for a class of cohomogeneity one metrics. This class includes, for example, the supersymmetric BMPV black hole \cite{Breckenridge:1996is}.

From the assumption of stationarity, we can define the mass as the Komar mass by integrating a two-form constructed from the time-like Killing vector $\partial_t$ over the three sphere at infinity as
\begin{equation}
\label{eqn:mass}
M = \frac{3}{32\pi G_5}\int_{S^3_{\infty}} \star_5 K
\end{equation}
where $K =dg$, and $g = g_{t \nu}d x^{\nu}$, and $G_5$ is Newton's constant in five dimensions. Similarly, from the Killing vectors $\partial_{\psi}$ and $\partial_{\phi}$ generating the rotation in the $\psi$ and $\phi$-planes respectively, we define the angular momentum by a similar Komar integral as
\begin{equation}
\label{eqn:rotation}
J =  \frac{1}{16\pi G_5}\int_{S^3_{\infty}} \star_5 K'
\end{equation}
where now $K' = dg'$, and $g' = g_{\mu \psi}dx^{\mu}$ or $g' = g_{\mu \phi}dx^{\mu}$. The electric charge we define as
\begin{equation}
\label{eqn:charge}
Q_{\rom{E}} = \frac{1}{16 \pi G_5} \int_{S^3_{\infty}} \bigg(\star_5 F  - \frac{1}{\sqrt{3}} F \wedge A\bigg).
\end{equation}
In our conventions, the BPS bound becomes
\begin{equation}
M \geq \sqrt{3}|Q_{\rom{E}}|.
\end{equation}

For completeness, let us also briefly recall the sigma-model machinery that we will use later in the paper. The Lagrangian (\ref{eqn:5dsugra}) of minimal supergravity when reduced over the orbits of two commuting Killing vectors $\partial_t$ and, say, $\partial_{\psi}$ gives rise to three dimensional Euclidean gravity on a manifold $\mathcal{M}_3$ coupled to a sigma-model of maps from $\mathcal{M}_3$ to the target space G$_{2(2)}/\mathrm{SO}_0(2,2)$. This sigma-model was derived and analysed in \cite{Bouchareb:2007ax, Compere:2009zh}. In this work we follow the conventions of  \cite{Compere:2009zh, Kim:2010bf}. To arrive at the three dimensional theory we use the reduction ansatz
\begin{align}
\mathrm{d}s^{2}_{5} &=
e^{\frac{1}{\sqrt3} \phi_{1}+\phi_{2}}d s^{2}_{3}
-e^{\frac{1}{\sqrt3} \phi_{1}-\phi_{2}}  (dt+\omega_3)^2
+e^{-\frac{2}{\sqrt3} \phi_{1}} ( d\psi +B_1 + \chi_1 dt)^{2},\label{eqn:canonicalform}
\\
\label{eqn:canonicalMaxwell}
A&=B_2+\chi_{3}dt + \chi_{2}d\psi.
\end{align}
%Note that the notation of the three dimensional one-forms differ somewhat from \cite{Compere:2009zh}, but conforms with the notation of \cite{Kim:2010bf}. We have also denoted the fifth coordinate as $\psi$ instead of  $z$ to emphasise that we work with five-dimensional Minkowski asymptotics and not with Kaluza-Klein asymptotics.

In terms of the fields introduced through this ansatz %(\ref{eqn:canonicalform})
we can rewrite the formulae for mass (\ref{eqn:mass}), angular momentum (\ref{eqn:rotation}), and charge (\ref{eqn:charge}) as
\be
M = \frac{3\pi}{16G_5}  \lim_{r\rightarrow \infty}\big(r^3\partial_{r}e^{\frac{1}{\sqrt3} \phi_{1}-\phi_{2}}\big), \quad  J_{\psi} = \frac{\pi}{8 G_5}  \lim_{r\rightarrow \infty}\big(r^3\partial_{r} (e^{-\frac{2}{\sqrt3} \phi_{1}}\chi_1)\big),  \quad Q_{\rom{E}} =\frac{\pi}{8G_5} \lim_{r\rightarrow \infty}\big(r^3\partial_{r} \chi_3\big), \label{charges}
\ee
where we have implicitly assumed the standard $1/r^2$ fall off for various metric and vector field components in asymptotically cartesian coordinates. In practice this simply means that the falloff of the scalars are such that limits  \eqref{charges} converge.

\subsection{Cveti\v c-Youm, BMPV, and extremal non-supersymmetric black holes}

A four parameter family of charged rotating black holes of five-dimensional minimal supergravity
was obtained by Cveti\v c and Youm in \cite{Cvetic:1996xz} by applying boosts and string dualities
to the (neutral, rotating) Myers-Perry black hole in five dimensions. The four parameters
specifying the solution can be chosen to be the mass $M$, two angular momenta $J_\phi$, $J_\psi$ and
electric charge $Q_\rom{E}$. It is convenient to present the solution in terms of the parameters
$m$, $l_1$, $l_2$, and $\delta$. We recall the full  solution in appendix \ref{app:CY}.

We restrict ourselves to the cohomogeneity one subclass of this solution, which is obtained by setting $l_1 = l_2$ in the form of the metric presented in appendix \ref{app:CY} and changing from the Hopf coordinates $(\tht, \phi_1, \phi_2)$ on the three-sphere to the Euler angular coordinates  $(\theta, \phi, \psi)$ introduced above. Setting $l_1=l_2=:l$ enhances the isometry group from $\mathbb{R}\times  \mathrm{U}(1) \times \mathrm{U}(1)$ to $\mathbb{R} \times \mathrm{SU}(2)\times  \mathrm{U}(1)$. This enhancement of the rotational symmetry allows us to describe the solution as a geodesic on the coset manifold. In terms of the parameters $m$, $l$, and $c = \cosh \delta, s = \sinh\delta$ the physical charges \eqref{charges} are
\be
M = \frac{3\pi}{4G_5}m(1+2s^2), \quad J_{\psi} = \frac{\pi}{4 G_5} m l (c^3+s^3), \quad J_{\phi}  = 0, \quad
\label{eqn:CYcharges} Q_{\rom{E}} = \frac{ \sqrt{3} \pi}{2G_5} m c s.
\ee
Writing the metric in the canonical form (\ref{eqn:canonicalform}) we also observe that the one-form $\omega_3$ in (\ref{eqn:canonicalform}) vanishes identically.

There are two regular extremal limits of this subclass of the Cveti\v c-Youm family. One of these limits is obtained by taking
$
m\to 0, l\rightarrow 0, \delta \to \infty$ keeping the mass, angular momentum and
charge finite.  In this limit we recover the supersymmetric BMPV black hole. The entropy of the BMPV black hole is given as
\be
S_\rom{BMPV} = 2 \pi \sqrt{N_\rom{M2}^3-4J_\psi^2}, \label{eq:SBMPV}
\ee
where $N_\rom{M2}= \frac{1}{\sqrt{3}} \ell_p Q_\rom{E}$ is the number of M2 branes, and $\ell_p$ is the five-dimensional planck length $\ell_p = \left(\frac{4G_5}{\pi}\right)^{1/3}$.
On the BMPV branch angular momentum $J_\psi$ can be continuously taken to zero from a finite non-zero value while maintaining regularity of the metric. In the limit $J_\psi =0$ the solution simply reduces to the supersymmetric 5D Reissner-Nordst\"om black hole. On this branch $ Q_{\rom{E}}^3$ must be greater than  $4 J_\psi^2$, as otherwise the solution develops naked closed timelike curves. The causal structure of the `over-rotating' BMPV spacetime has been discussed in thorough detail in \cite{Gibbons:1999uv}.

 The second extremal limit is obtained by taking $m = 2l^2$. This produces an extremal but non-supersymmetric black hole. Since the black hole is non-supersymmetric it does not saturates the BPS bound as can be easily checked from the expressions \eqref{eqn:CYcharges}. The entropy of the  black holes on this branch is given as
\be
S_\rom{extremal} = 2 \pi \sqrt{4J_\psi^2  -N_\rom{M2}^3}. \label{eq:SnonSUSY}
\ee
In this family the electric charge $Q_\rom{E}$ can be continuously taken to zero from a finite non-zero value while maintaining regularity of the metric. In the limit $Q_\rom{E} =0$ the solution reduces to extremal equally rotating Myers-Perry black hole. On this branch $N_\rom{M2}^3$ must be \emph{less} than  $4J_\psi^2$, otherwise the solution develops naked closed timelike curves.
%To the best of our knowledge, the causal structure of this `under-rotating' spacetime has not been discussed in detail in the literature. We study some of its properties in section \ref{sec:nilpotent}.

\section{Describing five dimensional black holes as geodesics}

\label{sec:geodesics}
%\subsection{Reductive decomposition}

In this section we discuss how to describe five-dimensional asymptotically flat black holes as geodesics. Let $\mathcal{V} : \mathcal{M}_3 \rightarrow \mathrm{G}_{2(2)}/\mathrm{SO}_0(2,2)$ be the coset map. If $\mathcal{V}$ only depend on the radial coordinate $r$ on $\mathcal{M}_3$ then the equations of motion for the sigma-model imply that $\mathcal{V}$ should trace out a geodesic on the coset manifold \cite{Breitenlohner:1987dg}. Define
$
\phi(g) =  (g^T\eta_4)^{-1} \eta_4
$ where $\eta_4$ is the symmetric matrix preserved by $\mathrm{SO}(2,2)$. We have that $\phi(g) = g$ if $g \in \mathrm{SO}(2,2)$. Using the map $\phi$ we define the group element
\begin{equation}
\label{eqn:nusquare}
M = \phi(\mathcal{V})^{-1} \mathcal{V}.
\end{equation}
A solution to the equations of motion is now given by
\begin{equation}
\label{eqn:geodesic}
M = M_0 \exp (\tau \mathcal{Q})
\end{equation}
provided $\phi(M) = M^{-1}$. Assuming the base space $\mathcal{M}_3$ to be flat,
\be
ds^2 = \frac{r^2}{4} \left[ dr^2 + \frac{1}{4} r^2 d \theta^2 + \frac{1}{4} r^2 \sin^2\theta d\phi^2 \right], \label{eqn:base}
\ee
Einstein's equations allow us to relate the affine parameter $\tau$ to the radial coordinate $r$. For an extremal five-dimensional black hole this gives
$\tau = -\frac{4}{r^2}.$
We interpret (\ref{eqn:geodesic}) as giving all geodesics from the point $M_0$. To describe asymptotically flat black holes we need to find the point $M_0$ from where the corresponding geodesics start, and solve the constraint $\phi(M) = M^{-1}$ at this point. This translates to finding a reductive decomposition
\begin{equation}
\label{eqn:5dreductive}
\mathfrak{g}_{2(2)} = \mathfrak{k}_5 \oplus \mathfrak{p}_5
\end{equation}
of the Lie algebra $\mathfrak{g}_{2(2)}$ such that $\mathfrak{p}_5 \cong T_{M_0}(G/\tilde{K})$ and $\mathfrak{k}_5$ is the Lie algebra of the group $K_5 \cong \mathrm{SO}_0(2,2)$ preserving the point $M_0$.

%\subsection*{Recalling four dimensional black holes}

To answer these questions, let us now recall how the similar construction work for asymptotically flat black holes in four dimensions. Assuming that the scalar fields in the sigma-model vanish at infinity, we have $M_0 = I$, where $I$ is the identity element in $\mathrm{G}_{2(2)}$. This implies that $\mathcal{Q} \in \mathfrak{g}_{2(2)}$ in (\ref{eqn:geodesic}) should obey
\begin{equation}
\label{eqn:cosetspace}
\mathcal{Q}^T \eta_4 - \eta_4 \mathcal{Q} = 0.
\end{equation}
Using $\eta_4$ we get the reductive decomposition
%\begin{equation}
%\label{eqn:4dreductive}
$\mathfrak{g}_{2(2)} = \mathfrak{k}_4 \oplus \mathfrak{p}_4$,
%\end{equation}
where $\mathfrak{k}_4 = \mathfrak{so}(2,2)$ is the Lie algebra of the $\mathrm{SO}(2,2)$ group preserving the identity and $\mathfrak{p}_4$ is defined as the space of elements obeying (\ref{eqn:cosetspace}), so that $T_I(G/\tilde{K}) \cong \mathfrak{p}_4$. Let us now investigate how all this translate to five-dimensional asymptotically flat black holes. %For vacuum gravity in five-dimensions this was worked out in
See also \cite{Giusto:2007fx}.

%\subsection*{Reductive decomposition}
%\label{sec:5dreductive}

To derive the asymptotic initial point $M_0$ we reduce five-dimensional Minkowski space (\ref{eqn:Minkowski}) to our G$_{2(2)}$ sigma model.
From the reduction ansatz (\ref{eqn:canonicalform}) we find
\be
\phi_1 = - \frac{\sqrt{3}}{2} \log \frac{r^2}{4} \nonumber ,\qquad
\phi_2 = - \frac{1}{2} \log \frac{r^2}{4}, \qquad
\chi_5 = \frac{r^2}{4},
\ee
and all other fields to be zero. Here $\chi_5$ is the scalar dual to $B_1$. This implies two things: (i) that five-dimensional Minkowski space is in fact described by a non-trivial geodesic on the coset manifold, contrary to the four dimensional case, and (ii) that the scalars diverge at infinity. However (\ref{eqn:nusquare}) constructed from the coset representative $\mathcal{V}_\rom{Mink}$ take a finite value when $r \rightarrow \infty$. Thus, we find, in the notation of \cite{Compere:2009zh}, that
 \begin{equation}
M_0 \equiv \lim_{r \rightarrow \infty} M_\rom{Mink} = \exp (- \tfrac{\pi}{12}(E_5-F_5)),
 \end{equation}
where $E_5$ and $F_5$ are Chevalley--Serre generators  of $\mathfrak{g}_{2(2)}$. %We can now rephrase the observation that the scalars diverge in the limit $r \rightarrow \infty$ as the fact that the compact geodesic
%\begin{equation}
%\gamma = \exp \tau(E_5-F_5)
%\end{equation}
%crosses the subspace of the manifold $G_{2(2)}$ where elements are not Iwasawa decomposable under $\mathrm{SO}(2,2)$ at the point $\tau = -\frac{\pi}{24}$. This situation is illustrated in figure %\ref{fig:cosetmanifold}.

Having found the point $M_0$ we can define a metric $\eta_5 := \eta_4 M_{0}$ such that $K_5$ is the Lie subgroup of $\mathrm{G}_{2(2)}$ that preserves this metric.  It is easy to see that for $g \in \mathrm{G}_{2(2)}$ with
$g^T \eta_5 g = \eta_5$ and if the coset representative transform as
$\mathcal{V} \rightarrow \mathcal{V}g, $
then the asymptotic point $M_0$ is preserved. Furthermore, if $M$ is given by (\ref{eqn:geodesic}), then the right action  $\mathcal{V} \rightarrow \mathcal{V}g, $ imply that
$
\mathcal{Q} \rightarrow Ad_g \mathcal{Q} .
$
So, the reductive decomposition (\ref{eqn:5dreductive}) is defined by elements satisfying
$\mathcal{Q}^T \eta_5 \pm \eta_5 \mathcal{Q} = 0$,  where we have $+$ for elements in $\mathfrak{k}_5$ and $-$ for elements in $\mathfrak{p}_5$, just as for the four dimensional decomposition. We conclude that asymptotically flat black holes map to elements $\mathcal{Q} \in \mathfrak{p}_5$, and under the action of the group preserving the asymptotics, these elements transform by conjugation. Exactly as in the four dimensional case we can hence classify different types of black hole solutions by the orbits of the group $K_5$. The space $\mathfrak{p}_5$ is eight-dimensional and spanned by the elements $Y_i$, $i=1,...,8$. Details about $Y_i$'s and the elements spanning $\mathfrak{k}_5$ are given in Appendix \ref{app:liealgebradetails}. The eight elements $Y_i$ correspond to  `conserved charges' (or more correctly to `conserved physical properties') of the black hole simply by the geodesic property. We now find the mapping between the eight elements of $\mathfrak{p}_5$ and physical properties of the five-dimensional spacetimes. These issues are also discussed  in \cite{Berkooz:2008rj}. See also \cite{Giusto:2007fx}.
%\subsection*{Asymptotic expansion and the sigma-model charges}
%\label{sec:charges}

We assume $\cQ^3 = 0$ and work with the three-dimensional base metric \eqref{eqn:base}. From the general nilpotent charge matrix $\cQ$ we construct the asymptotic spacetime fields and find the map from conserved quantities in the sigma-model to the conserved quantities in spacetime.
With a lot of hind-sight we parameterize the tangent vectors $\cQ$  in $\mathfrak{p}_5$ as
\begin{equation}
\label{eqn:chargematrix}
\mathcal{Q} = -N Y_1 + \beta_1 Y_2 + \frac{1}{\pi}G_5 Q_{\rom{E}} Y_3 -\frac{2}{\pi} G_5 J_{\psi} Y_4 + \frac{2\sqrt{3}}{\pi} G_5 J_\rom{M} Y_5 + \frac{1}{2  \pi \sqrt{3}} G_5 M Y_6 + \beta_2 Y_7 + p_0 Y_8 .
\end{equation}
Normalization of various coefficients here is chosen for later convenience. We now expand the fields, one at the time, in the reduction ansatz (\ref{eqn:canonicalform})  for large values of the radial coordinate $r$. The $\phi$-component of the one form  $\omega_3$ expands as
\begin{equation}
\omega_3|_{\phi} \simeq N\cos \theta + \mathcal{O}\left(\frac{1}{r^2}\right).
\end{equation}
This changes asymptotics of the spacetime. Therefore for asymptotic flatness we demand $N=0$. Expanding the three scalar fields $\chi_1, \chi_2$ and $\chi_3$ we find
\be
\chi_1 \simeq \frac{4N}{r^2} + \mathcal{O}\left(\frac{1}{r^4}\right), \quad
\chi_2 \simeq \frac{\beta_1}{p_0} + \mathcal{O}\left(\frac{1}{r^2}\right), \quad \mbox{and}
\ee
\be
\chi_3 \simeq
 \frac{1}{r^2}  \left(-\frac{4  \beta_1 ^2}{\sqrt{3} p_0}+\frac{4 N \beta_1}{p_0}-\frac{4 G_5 Q_\rom{E}}{\pi } \right) +\mathcal{O}\left(\frac{1}{r^4}\right).
%-4\frac{G_5 Q_{\rom{E}} p_0 +\sqrt{3} N \pi \beta_1+\sqrt{3} \pi \beta_1^2}{p_0 \pi r^2}+\mathcal{O}\left(\frac{1}{r^4}\right).
\ee
Note that parameter $\beta_1$ changes the asymptotic form of $\chi_3$; however, since it enters as a constant term in the expansion of $\chi_2$ it can be gauged away under certain natural assumptions on the nature of the spacetime. In the examples we consider $\beta_1$ always turns out to be zero. The $\phi$-component of the one form $B_1$ goes as
\begin{equation}
B_1|_{\phi} \simeq p_0 \cos\theta + \mathcal{O}\left(\frac{1}{r^2}\right).
\end{equation}
The parameter $p_0$ is thus the Chern class of the $\psi$-circle over the two-sphere parameterized by $(\theta, \phi)$. Therefore, $p_0$ is naturally interpreted as the orbifolding parameter, leading to constant $t, r$ hypersurfaces to be  $S^3/\mathbb{Z}_{p_0}$ for large values of $r$ \cite{Berkooz:2008rj}. For asymptotic flatness  we demand $p_0=1$. Let us therefore put $N = \beta_1 = 0$ and $p_0=1$ in the rest of the paper. The $\phi$-component of the one-form $B_2$ then expands as
\begin{equation}
B_2|_{\phi} \simeq \frac{4 \sqrt{3} G_5 J_\rom{M} \cos\theta}{ \pi r^2} + \mathcal{O}\left(\frac{1}{r^4}\right).
\end{equation}
This term is like a `dipolar magnetic flux' through the three-sphere. We will see later that for rotating electrically charged black holes this term is non-zero. Intuitively, it captures the angular momentum contained in the Maxwell field; for this reason we denote it by $J_\rom{M}$.  Finally, let us consider the two dilatons by expanding the $g_{tt}$ and $g_{t\psi}$ components. We find
\begin{equation}
g_{tt} \simeq - 1 + \frac{8G_5M}{3 \pi r^2}+ \mathcal{O}\left(\frac{1}{r^4}\right), \quad
g_{t\psi} \simeq - \frac{4G_5J_{\psi}}{\pi r^2} + \mathcal{O}\left(\frac{1}{r^4}\right).
\end{equation}
We conclude that $M$ refers to the mass, $Q_{\rom{E}}$ to the electric charge and $J_{\psi}$ to the angular momentum. Note that $\beta_2$ has so far not made an appearance. It can be thought of as  an auxiliary sigma-model charge.

Although, in the above analysis we have assumed $\cQ^3 = 0$ and have only worked with the three-dimensional base metric \eqref{eqn:base}, we expect the analysis to hold more generally. In particular, we will see in the next section that it applies to non-extremal black holes as well.

\section{Nilpotent orbits and phases of extremal black holes}
\label{sec:nilpotent}

In this section we consider nilpotent $\mathrm{SO}_0(2,2)$-orbits in $\mathfrak{p}_5$ and show that the partial ordering of nilpotent orbits of G$_{2(2)}$ is in one-to-one correspondence with the
phase diagram of extremal limits of the cohomogeneity one Cveti\v c-Youm black hole. The charge matrix of the non-extremal Cveti\v c-Youm black hole can be readily calculated, and it is given by
\be
\mathcal{Q}= \frac{\sqrt{3}}{2}  m c s Y_3 - \frac{1}{2}m l (c^3+s^3)Y_4 +  \frac{\sqrt{3}}{2} m l c s(c +s) Y_5 + \frac{\sqrt{3}}{8}  m(1+2  s^2)Y_6 + \frac{1}{16}m(m-8 l^2 )Y_7 +Y_8.
\ee
Note that the identification of charges derived in section \ref{sec:geodesics} matches with the charges given in section \ref{CY}. In particular $N = \beta_1 = 0$ and $p_0 = 1$. It can be easily verified that the charge matrix $\mathcal{Q} \in \mathfrak{p}_5$ obey the characteristic equation of \cite{Bossard:2009at}, namely
\begin{equation}
\mathcal{Q}^3-\frac{1}{4}\mathrm{tr}(\mathcal{Q}^2)\mathcal{Q} = 0,
\end{equation}
with
\begin{equation}
\label{eqn:trsquare}
\mathrm{tr}(\mathcal{Q}^2) = m (m-2l^2) .
\end{equation}
Note that the right hand side of this equation is independent of the electric charge parameter $\delta$. This is a reflection of the fact that the Cveti\v c-Youm black hole lies in the $\mathrm{SO}_0(2,2)$ orbit of the Myers-Perry black hole \cite{Bouchareb:2007ax, Compere:2009zh}. In fact, Cveti\v c and  Youm constructed their solution by applying boosts and string dualities which is equivalent to applying certain $\mathrm{SO}_0(2,2)$ transformations on the Myers-Perry black hole. The extremal limits $\mathrm{tr}(\mathcal{Q}^2) \rightarrow 0$ then give the nilpotent orbits of interest, as the zeros of equation (\ref{eqn:trsquare}) are exactly the two cases we discussed in section \ref{CY}. We now analyze these limits in more detail: we first recall the discussion of nilpotent orbits from \cite{Kim:2010bf} and then look at the two branches.

\begin{figure}
\begin{center}
\begin{overpic}[scale=0.4]{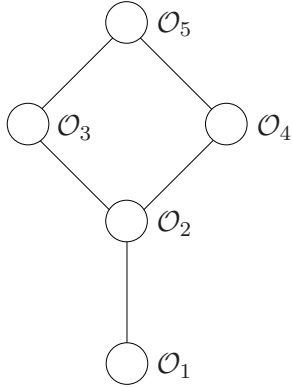}
\put(39,3){$\mathcal{O}_1$}
\put(39,40){$\mathcal{O}_2$}
\put(39,92){$\mathcal{O}_5$}
\put(65,65){$\mathcal{O}_4$}
\put(13,65){$\mathcal{O}_3$}
\end{overpic}
\caption{\small Hasse diagram for the partial ordering of the nilpotent orbits in $\mathfrak{g}_{2(2)}$.}
\label{fig:partialordering}
\end{center}
\end{figure}

The nilpotent orbits for the three dimensional sigma-model of minimal five-dimensional supergravity were analyzed in detail in \cite{Kim:2010bf}. The focus there was  on asymptotically Kaluza-Klein black holes, however, all of the group theoretical analysis carries over to the present  setting without any modification. Following \cite{Collingwood:1993}, reference \cite{Kim:2010bf} gave a detailed construction of representatives of all $\mathrm{SO}_0(2,2)$ orbits. % in the space $\mathfrak{p}_4$.
Via the reductive decomposition (\ref{eqn:5dreductive}) described in the previous section, we can immediately read off representatives for the nilpotent orbits of interest. We will not repeat the details of \cite{Kim:2010bf} here;  we simply remind the reader of the most important features. From \cite{Collingwood:1993} we know that nilpotent elements of $\mathfrak{g}_{2(2)}$ fit in five orbits $\mathcal{O}_i$, $i=1,...,5$ under the adjoint action of G$_{2(2)}$. These five orbits satisfy a partial ordering such that $\mathcal{O}_i$ is in the closure (in the usual topological sense) of some $\mathcal{O}_j$ with $i  < j$. In figure \ref{fig:partialordering} a Hasse diagram of the five nilpotent orbits of G$_{2(2)}$ is reproduced from \cite{Djokovic:2000}. The Hasse diagram encodes the fact $\mathcal{O}_3$ and $\mathcal{O}_4$ are in the closure of $\mathcal{O}_5$, and similarly that $\mathcal{O}_1$ is in the closure of $\mathcal{O}_2$, which in turn is in the closure of both $\mathcal{O}_3$ and $\mathcal{O}_4$. When restricting to $\mathrm{SO}_0(2,2)$ orbits of $\mathfrak{p}_4$ (or $\mathfrak{p}_5$), there is a possibility for these orbits to split into several SO$_0(2,2)$ orbits. This happens for $\mathcal{O}_3$ and $\mathcal{O}_4$ \cite{Kim:2010bf}. Both $\mathcal{O}_3$ and $\mathcal{O}_4$ `split' into two smaller orbits each in $\mathfrak{p}_4$ (or $\mathfrak{p}_5$). A similar splitting does not happen for  $\mathcal{O}_1$ and $\mathcal{O}_2$. The $\mathcal{O}_5$ orbit has not been analyzed in detail in the literature yet. The details of the $\mathcal{O}_5$ orbit does not concern us here, as the nilpotency degree of this orbit is too high to arise as the extremal limit of the Cveti\v c-Youm black holes. The four $\mathrm{SO}_0(2,2)$
 orbits for $\mathcal{O}_3$ and $\mathcal{O}_4$ are denoted \cite{Kim:2010bf}  $\mathcal{O}_{3K}, \mathcal{O}_{3K}', \mathcal{O}_{4K}$ and $\mathcal{O}_{4K}'$. It was shown in \cite{Kim:2010bf}, confirming the conjecture in \cite{Bossard:2009we}, that only one `splitted' orbit in each G$_{2(2)}$ orbit corresponded to physical black holes; the others  having singularities outside the horizon. We now show that a similar story holds for the five-dimensional asymptotically flat setting in minimal supergravity. It is natural to conjecture, following \cite{Bossard:2009we}, that a similar story holds for all five-dimensional theories.

Instead of following the method of \cite{Kim:2010bf}, where certain representatives of the orbits were used to generate the full orbit space, it turns out that we can solve the $\mathcal{Q}^3=0$ equation explicitly. This difference arises because in the analysis of \cite{Kim:2010bf} the only simplification made was setting the four dimensional NUT charge to zero, which is not enough for the equation $\mathcal{Q}^3=0$ to be solved explicitly, and as a result it was much more practical to generate orbits using the adjoint action of G$_{2(2)}$ on orbit representatives. In this paper we set  $\beta_1=0$, $N=0$ and $p_0=1$. This simplifies the problem considerably.  In fact, in the parametrization  (\ref{eqn:chargematrix}) of $\mathfrak{p}_5$ , we find that the solution space of the nilpotency equation $\mathcal{Q}^3=0$ has two distinct branches.  We now turn to the study of these branches.

%The equation $\mathcal{Q}^3=0$ translates to three equations so the resulting solution spaces become two dimensional.

\subsection{Supersymmetric branch} The first branch is 
\be
\beta_2 =  0, \quad J_\rom{M} =  J_\psi,  \quad M = \sqrt{3} Q_{\rom{E}}.
\label{eqn:susybranch}
\ee
We immediately see that this solution saturates the five-dimensional BPS bound. In fact the solution is supersymmetric. The condition that the mass be positive translates to the condition that electric charge must be positive.
The space-time solution derived from the geodesic is exactly the BMPV solution. Thus, as expected,  (\ref{eqn:susybranch}) reproduces the $m \rightarrow 0,~l\rightarrow 0,~\delta \rightarrow \infty$ limit of the Cveti\v c-Youm black hole. As is well known, for the BMPV solution to have all closed time-like curves hidden behind the horizon, we need
\begin{equation}
\label{eqn:physicalboundBMPV}
N_\rom{M2}^3> 4J_\psi^2.
\end{equation}
In our way of seeing it, this is equivalent to the condition that the exponentials in (\ref{eqn:canonicalform}) are always positive. In fact, for BMPV we have
\begin{equation}
\label{eqn:psicomponent}
\lim_{r\rightarrow 0}e^{-\frac{2}{\sqrt3} \phi_{1}} = \left(\frac{G_5}{2\pi}\right)^{2/3} \frac{N_\rom{M2}^3-4J_\psi^2}{N_\rom{M2}^2}  ,
\end{equation}
so for the exponentials to be well-defined we need to satisfy (\ref{eqn:physicalboundBMPV}). Group theoretically this bound is equivalent to the border between the two SO$_0(2,2)$ orbits $\mathcal{O}_{3K}$ and $\mathcal{O}_{4K}$. $\mathcal{O}_{4K}$ contain the over-rotating BMPV spacetime and $\mathcal{O}_{3K}$ the under-rotating. That (\ref{eqn:physicalboundBMPV}) is violated also signals closed time-like curves outside of the horizon, since (\ref{eqn:psicomponent}) is the $g_{\psi\psi}$-component of the metric for the black hole. See \cite{Gibbons:1999uv} for a more detailed discussion about this point. One also sees that the entropy
\begin{equation}
\label{eqn:horizonBMPV}
S = 2\pi \sqrt{N_\rom{M2}^3-4J_\psi^2},
\end{equation}
is ill-defined unless we obey (\ref{eqn:physicalboundBMPV}). When the bound (\ref{eqn:physicalboundBMPV}) is saturated we are in the orbit $\mathcal{O}_2$, a zero-horizon black hole with $ 4J_\psi^2 = N_\rom{M2}^{3}$. Furthermore, if all charges vanish we end up in  the $\mathcal{O}_1$ orbit, where the only solution is Minkowski space. 

\subsection{Non-supersymmetric branch}
\begin{figure}
\begin{center}
\begin{overpic}[scale=0.8]{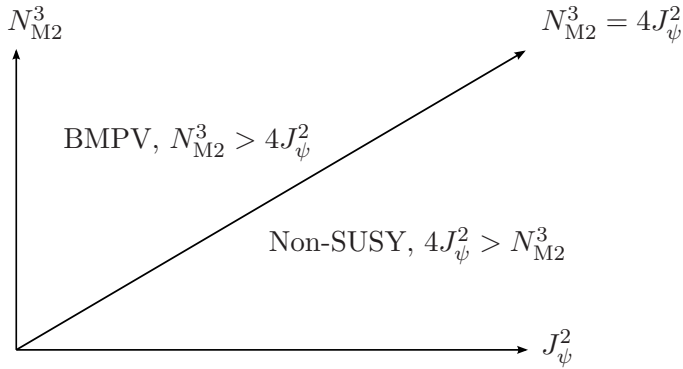}
\put(-1,63){$N_{\rom{M2}}^3$}
\put(103, 0){$J_\psi^2$}
\put(103, 63){$N_{\rom{M2}}^3=4J_\psi^2$}
\put(10, 40){BMPV, $N_{\rom{M2}}^3>4J_\psi^2$}
\put(50,20){Non-SUSY, $4J_\psi^2 > N_{\rom{M2}}^3$}
\end{overpic}
\caption{\small A phase diagram of the extremal limits of the cohomogeneity one  Cveti\v c-Youm black hole. The line $N_{\rom{M2}}^3=4J_\psi^2$ corresponds to the $\mathcal{O}_2$ orbit, the region above the line   corresponds to the $\mathcal{O}_{3K}$ orbit and the region below the line corresponds to $\mathcal{O}_{4K}'$ orbit. The origin being Minkowski space corresponds to $\mathcal{O}_1$. Compare this phase diagram with the Hasse diagram in figure \ref{fig:partialordering}.}
\label{fig:phasediagram}
\end{center}
\end{figure}

The other branch is given by
\be
M  = \frac{18 \pi}{G_5} \frac{J_\rom{M}^2}{Q_\rom{E}^2} - \sqrt{3} Q_\rom{E} , \quad
J_\psi =\frac{12\sqrt{3} \pi}{G_5}\frac{J_\rom{M}^3}{Q_\rom{E}^3} - 3 J_\rom{M}, \quad
\beta_2 =  \frac{12\sqrt{3} G_5}{\pi}\frac{J_\rom{M}^2}{Q_\rom{E}} - 108\frac{J_\rom{M}^4}{Q_\rom{E}^4}.
\label{eqn:nonsusycharges}
\ee
Here, to satisfy the BPS bound, we have to obey the condition
\begin{equation}
\label{eqn:physicalboundNonSUSY}
4 J_\psi^2 > N_\rom{M2}^3.
\end{equation}
This branch corresponds to the $m=2l^2$ extremal limit of the cohomogeneity one Cveti\v c-Youm black hole. If we consider the positivity of the exponentials in the solution, we find that the condition (\ref{eqn:physicalboundNonSUSY}) is sufficient for the metric to be well-defined. Exactly opposite to the supersymmetric branches above, the physical orbit $\mathcal{O}_{4K}'$ now contain the over-rotating black hole, and the unphysical orbit $\mathcal{O}_{3K}'$ contain the under-rotating spacetime. The under-rotating spacetime violates the BPS-bound. The regular solution on this branch is analyzed in detail in \cite{Dias:2007nj}. We note that $g_{\psi\psi}$ becomes negative at finite positive $r$ if $4J_\psi^2 < N_\rom{M2}^3$, making closed orbits of $\partial_{\psi}$ time-like. In figure \ref{fig:phasediagram} we illustrate the phase space of physical solutions in terms of the charges $N_\rom{M2}$ and $J_\psi$. Note the similarities between this phase diagram and the Hasse diagram of the partial ordering in figure \ref{fig:partialordering}.
 %One can also check that $\partial_{\psi}$ always becomes null outside of the ergo-sphere if the bound (\ref{eqn:physicalboundNonSUSY}) is violated.

\section{Discussion}
\label{sec:disc}
In this paper we studied extremal cohomogeneity one five-dimensional asymptotically flat black holes of minimal supergravity in terms of the geodesics generated  by nilpotent elements of the Lie algebra $\mf{g}_{2(2)}$ on the coset manifold $\mathrm{G}_{2(2)}/\mathrm{SO}(2,2)$. We discussed two branches of regular extremal black holes with these properties: (i) the supersymmetric BMPV branch, and (ii) the non-supersymmetric extremal branch. We showed that these branches are reproduced by $\cO_{3K}$ and $\cO_{4K}'$ nilpotent $\mathrm{SO}(2,2)$-orbits respectively. We also showed that the partial ordering of nilpotent orbits of $\mathrm{G}_{2(2)}$ is in one-to-one correspondence with the phase diagram of these extremal black holes.

The  correspondence between  phase diagrams of extremal four and five-dimensional black holes and the partial ordering of nilpotent orbits of the corresponding three-dimensional hidden symmetry group seems to be a generic phenomenon. So far this has only been explored for  cohomogeneity one metrics. It is likely that these considerations extend beyond this assumption. For example extremal rotating Kaluza-Klein black holes (D0-D6) of vacuum five-dimensional gravity come in two branches: (i) slowly rotating branch, and (ii) fast rotating branch. The entropy for the slowly rotating solutions take the form
\be
S_\rom{slow} = \pi \sqrt{N_{0}^2 N_{6}^2 - 4 J^2}
\ee
and  for the fast rotating solutions take the form
\be
S_\rom{fast} = \pi \sqrt{4 J^2 - N_{0}^2 N_{6}^2}.
\ee
Given the similarity of these expressions with equations \eqref{eq:SBMPV} and \eqref{eq:SnonSUSY} it may be true that the consideration of nilpotent orbits extend to cohomogeneity two attractors as well. We hope to return to this question in the future.

\subsection*{Acknowledgements}
We thank Boris Pioline for encouraging us to explore the five-dimensional asymptotically flat boundary conditions. We have benefited from our discussions with Guillaume Bossard, Stefano Giusto, and Clement Ruef. Our work is partially supported by IISN - Belgium (conventions 4.4511.06
and 4.4514.08) and by the Belgian Federal Science Policy Office through the
Interuniversity Attraction Pole P6/11.

\appendix

\section{Cveti\v c-Youm metric}
\label{app:CY}
For completeness we present the Cveti\v c-Youm solution in this appendix. See also \cite{Cvetic:1997uw, Cvetic:1996xz, Giusto:2004id,  Jejjala:2005yu, Dias:2007nj, Figueras:2009mc}. The metric and the vector potential take the form
\be
ds^2 = g_{tt}\, dt^2 + 2g_{t\phi_1}\, dtd\phi_1 + 2g_{t\phi_2}\, dtd\phi_2
     + g_{\phi_1\phi_1}\, d\phi_1^2 + g_{\phi_2\phi_2}\, d\phi_2^2 + 2g_{\phi_1\phi_2}\, d\phi_1d\phi_2
     + g_{rr}\, dr^2 + g_{\tht\tht}\, d\tht^2  \ ,
\label{metricCY}
\ee
\be
A = A_{t} dt + A_{\phi_1} d \phi_1 + A_{\phi_2} d\phi_2 \ ,
\ee
with
\bea
g_{tt}       &=&  -\, \frac{\Sg(\Sg - 2m)}{(\Sg + 2ms^2)^2}  \ , \nn  \\
g_{t\phi_1}    &=&  -\, \frac{2m \sin^2\tht \left[ \Sg \left\{ l_1c^3 + l_2s^3 \right\} -2ml_2s^3\right] }{(\Sg + 2ms^2)^2}  \ , \nn \\
g_{t\phi_2}    &=&  -\, \frac{2m \cos^2\tht \left[ \Sg \left\{ l_2c^3 + l_1s^3
\right\} - 2ml_1s^3\right] }{(\Sg + 2ms^2)^2}  \ , \nn
\eea
\bea
g_{\phi_1\phi_1} &=&  \frac{\sin^2\tht}{(\Sg + 2ms^2)^2} \left[ (r^2+2ms^2+l_1^2)(\Sg + 2ms^2)^2 \right. \nn \\
  &+& \left. 2m\sin^2\tht \left\{ \Sg(l_1^2c^2-l_2^2s^2)
  - 4ml_1 l_2 c^3s^3 - 2ms^4(l_1^2c^2+l_2^2s^2) - 4m l_2^2 s^4 \right\} \right]  \ , \nn \\
g_{\phi_2\phi_2} &=&  \frac{\cos^2\tht}{(\Sg + 2ms^2)^2} \left[ (r^2+2ms^2+l_2^2)(\Sg + 2ms^2)^2 \right. \nn \\
  &+& \left. 2m\cos^2\tht \left\{ \Sg(l_2^2c^2-l_1^2s^2)
  - 4ml_1 l_2 c^3s^3 - 2ms^4(l_2^2c^2+l_1^2s^2) - 4m l_1^2 s^4 \right\} \right]  \ , \nn \\
g_{\phi_1\phi_2} &=&  \frac{2m \cos^2\tht \sin^2\tht \left[ l_1 l_2 \left\{ \Sg-6ms^4 \right\} - 2m(l_1^2+l_2^2)s^3c^3
                - 4ml_1 l_2 s^6 \right] }{(\Sg + 2ms^2)^2}  \ , \nn
\eea
\bea
g_{rr}       &=&  \frac{r^2(\Sg + 2ms^2)}{(r^2+l_1^2)(r^2+l_2^2)-2mr^2}  \ , \nn \\
g_{\tht\tht} &=&  \Sg + 2ms^2  \ , \nn \\
A_{t} &=& -\frac{2 \sqrt{3} m s c}{(\Sg + 2ms^2)} \nn \ , \\
A_{\phi_1} &=& - A_t  (l_1 c + l_2 s) \sin^2 \tht \nn \ , \\
A_{\phi_2} &=& - A_t  (l_2 c + l_1 s)\cos^2 \tht  \ .
\eea
For convenience we have defined
\be
\Sg(r,\tht)  \equiv  r^2 + l_1^2 \cos^2\tht + l_2^2 \sin^2\tht   \ .
\ee
Setting $\delta=0$ reproduces the five-dimensional MP black hole. In the above form Hopf coordinates are used on the $S^3$. In the main text we use Euler coordinates on the $S^3$. The two coordinate systems are related as
\be
\tht = \frac{\theta}{2}, \quad  \psi = \phi_1 + \phi_2, \quad \phi = \phi_2 - \phi_1.
\ee
For calculational convenience we also use $x = \cos \tht$. Our spacetime orientation is \be \epsilon_{rx\phi_1 \phi_2 t} =  r \cos \tht \left(  \Sg + 2 m s^2\right).\ee

%\numberwithin{equation}{section}

\section{Asymptotic reductive decomposition}
\label{app:liealgebradetails}
In this appendix we list the explicit form of $\mathfrak{k}_5$ and $\mathfrak{p}_5$ in the reductive decomposition
\begin{equation}
\mathfrak{g}_{2(2)} = \mathfrak{k}_5 \oplus \mathfrak{p}_5
\end{equation}
at the point $M_0 \in \mathrm{G}_{2(2)}/\mathrm{SO}_0(2,2)$ derived in the main text. The subalgebra $\mathfrak{k}_5$ corresponding to $K_5$ is
\begin{equation}
\mathfrak{k_5} = \mathrm{Span}_{\mathbb{R}}(X_1, X_2,X_3,X_4,X_5,X_6)
\end{equation}
with
\begin{align}
X_1 &= e_1-f_6, \quad
X_2  = f_4-f_2, \quad
X_3  = e_3+f_3, \\
X_4  &=  f_1-e_6, \quad
X_5  =  e_2-e_4, \quad
X_6  =  h_1 + \frac{1}{\sqrt{3}}h_2,
\end{align}
where we follow the notation of \cite{Compere:2009zh} for the definitions of $h$'s, $e$'s, and $f$'s. This subalgebra $\mathfrak{k}_5$ is isomorphic to $\mathfrak{sl}(2, \mathbb{R}) \oplus \mathfrak{sl}(2, \mathbb{R})$ as can be seen via the following change of basis
\begin{align}
\mathbf{h}_1 &= \frac{\sqrt{3}}{2} (X_3+\frac{3}{2} X_6), \quad
\mathbf{e}_1 =3(X_4-\frac{1}{\sqrt{3}} X_5), \quad
\mathbf{f}_1 = \frac{3}{8}(X_1+\frac{1}{\sqrt{3}} X_2),\\
\mathbf{h}_2 &= -\frac{\sqrt{3}}{2} (X_3-\frac{1}{2} X_6), \quad
\mathbf{e}_2 =X_4+\sqrt{3} X_5, \quad
\mathbf{f}_2 = \frac{1}{8}(X_1-\sqrt{3} X_2).
\end{align}
The elements in $\mathfrak{g}_{2(2)}$ that span $\mathfrak{p}_5$ are
%\begin{equation}
%\mathfrak{p}_5 = \mathrm{Span}_{\mathbb{R}}(Y_1, Y_2, Y_3, Y_4, Y_5, Y_6, Y_7, Y_8)
%\end{equation}
%with
\begin{align}
Y_1 &=e_1+f_6, \quad
Y_2 =f_4+f_2, \quad
Y_3 =e_3-f_3, \quad
Y_4 =f_1+e_6, \\
Y_5 &=e_2+e_4, \quad
Y_6 =h_1-\sqrt{3}h_2, \quad
Y_7 =e_5, \quad
Y_8 =f_5.
\end{align}
%We see that
These eight generators correspond to eight conserved charges in the three-dimensional sigma-model. %The question is what five dimensional charges they correspond to.
Let us now diagonalize the adjoint action of $\bf h_1$ and $\bf h_2$ on $\mathfrak{p}_5$. By doing so we can immediately read of the representatives of nilpotent orbits from the discussion in \cite{Kim:2010bf}. Let $Y_{(m,n)}$ be the generators of $\mathfrak{p}_5$ with eigenvalues $m$ under $\bf h_1$ and $n$ under $\bf h_2$, then
%\begin{eqnarray}
\begin{align}
Y_{(3,1)} & = Y_7, \quad
Y_{(1,1)} =  Y_4+\frac{Y_5}{\sqrt{3}}, \quad
Y_{(-1,1)} = Y_3-\frac{Y_6}{2}, \quad
Y_{(-3,1)} =Y_1-\sqrt{3}Y_2, \\
Y_{(3,-1)}& = Y_4-\sqrt{3}Y_5, \quad
Y_{(1,-1)} = Y_3+\frac{Y_6}{2}, \quad
Y_{(-1,-1)} = Y_1+\frac{Y_2}{\sqrt{3}}, \quad
Y_{(-3,-1)} =  Y_8.
%\end{eqnarray}
\end{align}
%\subsection*{Truncation to pure gravity}
The truncation of the sigma-model to the pure gravity subalgebra $\mf{sl}(3, \mathbb{R})$ is   obtained by picking out the long roots. The isotropy subgroup for pure gravity $\mathfrak{so}(2,1) \cong \mathfrak{sl}(2, \mathbb{R})$ is spanned by
\begin{equation}
H = \sqrt{3} X_6, \quad E = 2 X_4, \quad F = X_1,
\end{equation}
and the coset space is spanned by $Y_1, Y_4, Y_6, Y_7$ and $Y_8$.

\bibliographystyle{utphys}
\bibliography{Refdata}

\end{document}